\documentclass{icrc2009}

\usepackage{graphicx}   
\usepackage[caption=false]{caption}    
\usepackage[font=footnotesize]{subfig} 
\usepackage{fixltx2e}
\usepackage{url}

\newcommand{\shorttitle}[1]%
{\markboth{Proceedings of the 31\MakeLowercase{$^{st}$} ICRC, {\L}\'{o}d\'{z} 2009}{#1} }
\newcommand{\etal}{\MakeLowercase{\textit{et al. }}} 

\def\degr{\hbox{$^\circ$}}
\def\arcmin{\hbox{$^\prime$}}
\def\arcsec{\hbox{$^{\prime\prime}$}}

\begin{document}
\title{Detection of very-high-energy $\gamma$-ray emission from the
  vicinity of PSR\,B1706$-$44 with H.E.S.S.}

\author{\IEEEauthorblockN{S. Hoppe\IEEEauthorrefmark{1}, E. de O\~{n}a
    Wilhelmi\IEEEauthorrefmark{1}, B. Kh\'elifi\IEEEauthorrefmark{2},
    R. C. G. Chaves\IEEEauthorrefmark{1},\\
    O.C. de Jager\IEEEauthorrefmark{3}, C. Stegmann\IEEEauthorrefmark{4}
    and R. Terrier\IEEEauthorrefmark{5} for the H.E.S.S. Collaboration}
  \\
  \IEEEauthorblockA{\IEEEauthorrefmark{1} Max-Planck-Institut f\"ur
    Kernphysik, Heidelberg, Germany}
  \IEEEauthorblockA{\IEEEauthorrefmark{2} Laboratoire
    Leprince-Ringuet, Ecole Polytechnique, CNRS/IN2P3, Palaiseau,
    France} \IEEEauthorblockA{\IEEEauthorrefmark{3} Unit for Space
    Physics, North-West University, Potchefstroom, South Africa}
  \IEEEauthorblockA{\IEEEauthorrefmark{4} Universit\"at
    Erlangen-N\"urnberg, Physikalisches Institut, Germany}
  \IEEEauthorblockA{\IEEEauthorrefmark{5} Astroparticule et Cosmologie
    (APC), CNRS, Universit\'e Paris VII, Paris, France}}

\shorttitle{S. Hoppe \etal (H.E.S.S Collaboration) Detection of HESS~J1708$-$443}
\maketitle

\begin{abstract}
  The energetic pulsar PSR\,B1706$-$44 and the adjacent
  supernova remnant (SNR) candidate G\,343.1$-$2.3 were observed by H.E.S.S. during a dedicated observational
  campaign in 2007.  A new source
  of very-high-energy (VHE; E\,$>$\,100\,GeV) $\gamma$-ray emission, HESS\,J1708--443,
  was discovered with its centroid at
  $\alpha_{2000}$\,$=$\,17$^{h}$8$^{m}$10$^{s}$ and
  $\delta_{2000}$\,$=$\,$-$44\degr21\arcmin~($\pm$\,3\arcmin$_{\mathrm{stat}}$~on each axis)
  The VHE
  $\gamma$-ray source is significantly more extended than the
  H.E.S.S. point-spread function, with an intrinsic Gaussian width of
  0.29\degr\,$\pm$\,0.04\degr. Its energy spectrum can be described by a
  power law with a photon index 
  $\Gamma$\,$=$\,2.0\,$\pm$\,0.1$_{\mathrm{stat}}\,\pm$\,0.2$_{\mathrm{sys}}$.
  The integral flux measured between 1$-$10\,TeV
  is $\sim$\,17\% of the Crab Nebula flux in the same energy range. The possible associations with
  PSR\,B1706--44 and SNR G\,343.1--2.3 are discussed.
  \end{abstract}


\section{Introduction}
The pulsar PSR\,B1706$-$44 was first detected in a high-frequency radio
survey \cite{Johnston_1992}. With a spin period of
102\,ms, a characteristic age of 17\,500 yr and a spin-down
luminosity of 3.4\,$\cdot$\,10$^{36}$\,erg\,s$^{-1}$, it belongs to the
class of relatively young and very energetic pulsars. Estimates for its 
distance range from 1.8\,kpc \cite{Johnston_1992} \cite{Taylor_1993} to
3.2\,kpc \cite{Koribalski_1995}. The positionally-coincident
$\gamma$-ray source 2CG342$-$02 \cite{Swanenburg_1981} was firmly
identified with PSR\,B1706$-$44 when EGRET observed
pulsed emission with the same period seen in the radio waveband
\cite{Thompson_1992}. PSR\,B1706$-$44 is therefore one of the very
first pulsars from which pulsed emission was detected not only in radio
\cite{Johnston_1992} and X-rays \cite{Gotthelf_2002}, but also in
high-energy $\gamma$-rays.

The pulsar PSB~B1706$-$442 is surrounded by a synchrotron nebula with an
extension of ~3\arcmin~at radio wavelengths
\cite{Frail_1994} \cite{Giacani_2001}. The observed polarization and the
flat spectrum of the radio emission (photon index of 0.3)
suggest a pulsar wind nebula (PWN) origin. The synchrotron nebula is
also visible in X-rays, first reported by Finley et
al.~\cite{Finley_1998} using \emph{ROSAT} observations. Employing the
superior resolution of \emph{Chandra}, Romani~et~al.~\cite{Romani_2005}
were able to map the morphology of the PWN at
the arcminute scale. Their findings suggest a diffuse PWN with a
spectral index of 1.77, surrounding a more complex structure
comprising a torus and inner and outer jets. The diffuse PWN has a
radius of $\sim$110\arcsec~and exhibits a fainter, longer extension
to the West. The non-deformed X-ray jets support the low scintillation
velocity of the pulsar of less than 100\,km\,s$^{-1}$~\cite{Johnston_1998}.

PSR\,B1706$-$44 is located at the southeast end of an incomplete arc of
radio emission \cite{McAdam_1993} suggested to be the shell of a faint
supernova remnant (SNR G\,343.1$-$2.3). The arc itself is embedded
in weak, broad-scale radio emission \cite{Frail_1994} for which
polarization measurements suggest an association with synchrotron
radiation from the SNR \cite{Dodson_2002}. No X-ray emission was
detected from the radio structure (see e.g.~\cite{Becker_1995}). The
question of a possible association between PSR\,B1706$-$44 and
G\,343.1$-$2.3 could not be answered unambiguously so far. The
dispersion distance for the pulsar of 2.3\,$\pm$\,0.3\,kpc
(\cite{Romani_2005} and references therein) using the free electron
distribution model by Cordes and Lazio~\cite{Cordes_2002} is
compatible with the $\Sigma-D$ distance of $\sim$3\,kpc for the SNR
\cite{McAdam_1993}. However, the off-center position of the pulsar
relative to the radio-arc implies a rather high proper motion velocity
($\sim$700\,km\,s$^{-1}$) which is incompatible with the measured
scintillation velocity. Bock et~al.~\cite{Bock_2002} suggested a
scenario whereby an off-centered cavity explosion would release the
restrictions on the implied velocity and invalidate the age estimate
for the SNR of $\sim$5\,000\,yr \cite{McAdam_1993}, which is based on a
Sedov-Taylor model. In this scenario, the radio arc is identified with
the former boundary of the wind-blown cavity that was overtaken and
compressed by the expanding SNR. The diffuse, broad-scale radio
emission would then result from the interaction of the SNR with the
parent molecular cloud.

At very-high energies (VHE; E\,$>$\,100 GeV), the region of interest
was observed using ground-based atmospheric Cherenkov telescopes. The
CANGAROO Collaboration reported the detection of steady emission
coincident with the pulsar using the 3.8-m CANGAROO-I telescope in 1992--1993 \cite{Ogio1993} \cite{Kifune1995}.
They measured an integral flux above 1~TeV of $\sim$35\% 
of the Crab Nebula flux.  It was later revealed that the actual mirror
reflectivity at the time of the observations would have resulted in a higher
minimum energy threshold of $\sim$2~TeV \cite{Roberts1997}.
The 4-m BIGRAT telescope \cite{Rowell1998} also observed the pulsar in 1993--1994 and reported a compatible 
upper limit (UL).
Observations in 1996 with the Durham Mark 6 telescope \cite{Chadwick1998} appeared to confirm
the detection, with a reported integral flux that was compatible within the
large systematic uncertainties ($\pm$\,30\% for CANGAROO-I and $\pm$\,50\% for the Mark 6).  
Further observations with the 10-m CANGAROO-II telescopes in 2000--2001 again seemed to confirm
the detection.  However, when the H.E.S.S. Collaboration observed the pulsar in 2003 during
its commissioning phase---operating only two out of four telescopes, without a stereo hardware trigger---they
did not detect any significant VHE $\gamma$-ray emission from the vicinity of PSR\,B1706--44.
The derived UL on the integral flux was found to be 
$\sim$5\% of the Crab, in stark disagreement with the
previous findings \cite{Aharonian_2005_1706}.  Shortly thereafter, preliminary analysis
of stereo observations with the 4\,$\times$\,10-m CANGAROO-III telescope array 
did not confirm the earlier CANGAROO-I detection but instead resulted in
an UL of $\sim$10\% Crab \cite{Tanimori2005}, in agreement with the H.E.S.S. results.
Very recently, the CANGAROO Coll.~undertook a comprehensive re-analysis of their archival
CANGAROO-I data and now find an UL to the integral flux at $\sim$13\% Crab \cite{Yoshikoshi2009}, also
compatible with the H.E.S.S. UL.  In 2007, additional H.E.S.S. data was taken on the
pulsar, now utilizing the superior sensitivity of the fully-operational H.E.S.S. telescope array. 
In this proceeding, the findings of this observation campaign are presented. No point-like emission is
detected at the pulsar position. However, an extended source of VHE
$\gamma$-rays was discovered in the region of interest. Its centroid
appears significantly displaced from the pulsar position. Although the
measured flux from the extended region exceeds the previously-published UL
by a small margin, a re-analysis of the older
H.E.S.S. data set (originally published in
\cite{Aharonian_2005_1706}), using the up-to-date H.E.S.S. standard
analysis framework, yields revised flux ULs which are consistent
with the currently-detected flux.

\begin{figure}[t!]
\centering
   \includegraphics[width=3.0in]{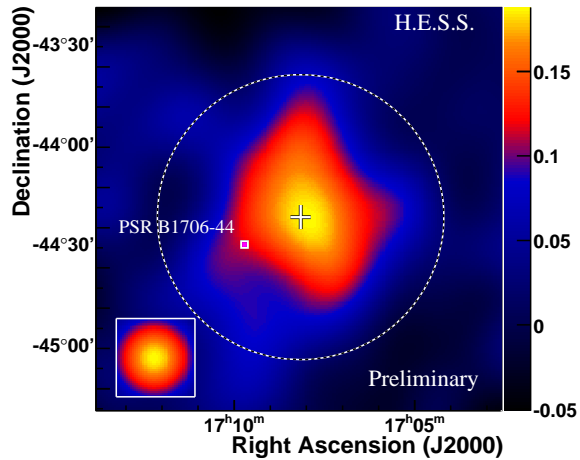}
   \caption{Image of the VHE $\gamma$-ray excess from
     HESS\,J1708$-$443, smoothed with a Gaussian profile of
     $\sigma$\,$=$\,0.15\degr along each axis. The white cross indicates the
     best fit position of the center-of-gravity of the emission
     together with its statistical errors. The white circle illustrate
     the region which was used for spectral analysis. The position of
     the pulsar PSR\,B1706$-$44 is marked by a square. The inset in the
     bottom left corner shows the point-spread function of the
     instrument for this particular data set smoothed in the same way
     as the excess map.}
\label{fig1}
\end{figure}
\section{The H.E.S.S. telescopes / analysis technique}
The High Energy Stereoscopic System (H.E.S.S.) is an array of four,
imaging atmospheric Cherenkov telescopes, dedicated to the observation
of VHE $\gamma$-rays. The array is located in the Khomas Highlands of
Namibia (23\degr16'17''\,S, 16\degr29'58''\,E). Each telescope is
equipped with a tessellated, spherical mirror of 107\,m$^{2}$ area and
a camera comprised of 960 photomultiplier tubes, covering a
field-of-view (FoV) 5\degr~in diameter. The telescopes are operated
in coincidence mode, which requires a trigger of at least two
telescopes for an air shower to be recorded. The stereoscopic approach
allows a high angular resolution of $<$\,0.1\degr\space per event,
a good energy resolution of $\sim$16\% (on average) and an effective
background rejection \cite{Aharonian_2006_Crab}. The H.E.S.S. array
can detect point sources at flux levels of about 1\% of the Crab
Nebula flux near zenith with a statistical significance of 5\,$\sigma$
in 25\,h of observations. Its large FoV and good off-axis sensitivity not only make it
ideally suited for surveying the Galactic plane \cite{AharonianGPS1} \cite{AharonianGPS2}
\cite{Chaves2009}, but also for studying extended sources like HESS~J1708--443.

The region of interest, which includes PSR\,B1706$-$44 and the SNR
G\,343.1$-$2.3, was observed with the full H.E.S.S. telescope array in
2007. The observations were dedicated to search for VHE $\gamma$-ray
emission from the pulsar and were therefore taken in \emph{wobble} mode,
alternating around its radio position
($\alpha_{2000}$\,=\,17$^{h}$9$^{m}$42.73$^{s}$,
$\delta_{2000}$\,=\,$-$44\degr29\arcmin8.2\arcsec~\cite{Wang_2000}). In
this observation mode, the array is pointed towards a position offset
from the source of interest to allow for simultaneous background
estimation.

The data set was analyzed using the Hillas second-moment method \cite{Aharonian_2006_Crab}.
For $\gamma$-hadron separation, {\it hard cuts} were used,
which require a minimum of 200 photo electrons (p.e.) to be recorded per
shower image. Compared to {\it std cuts} (80 p.e.), this relatively
strict requirement results in better background rejection and an
improved angular resolution, but also in an increased energy
threshold (560\,GeV for this data set). The time-dependent optical
response of the system was estimated from the Cherenkov light of
single muons passing close to the telescopes \cite{Bolz_2004}.

Three different background estimation procedures
\cite{Berge2007} were used in this analysis. For 2D image
generation, the Ring Background Method was used with a mean ring radius
of 0.85\degr. Since this method includes an energy-averaged model for the
camera acceptance to account for the different offsets of the signal and
background regions from the camera center, it was not used for
spectral extraction. The Reflected Region Method was instead used
to measure the flux from the pulsar position. For spectral
extraction from very extended regions which also enclose the pointing
positions of the telescopes, the background was estimated from
off-source (OFF) data taken in regions of the sky where no $\gamma$-ray
sources are known. To match the observing conditions between on-source (ON)
and OFF data, the two observations had to be taken within
six months of each other and at similar zenith angles, in a procedure similar
to that used for Vela~Jr \cite{Aharonian2005VelaJr}. The normalization
between ON and OFF observations was performed using the total
event number in the two observations, excluding regions with significant VHE
$\gamma$-ray signal.

\section{Results}
Figure \ref{fig1} shows the excess count map of the
2\degr\,$\times$\,2\degr\space region around the source smoothed with
a Gaussian profile of width 0.15\degr\space to reduce statistical
fluctuations. A clear excess of VHE $\gamma$-rays is observed with a
peak statistical significance of 7.5 $\sigma$ using an integration
radius of $\theta$\,$=$\,0.4\degr. Fitting the fine-binned and
unsmoothed excess map with a radially-symmetric Gaussian profile
($\phi$\,$=$\,$\phi_0e^{-r^{2}/(2\sigma^{2})}$) convolved with the
point-spread function (PSF) of the instrument leads to a best fit position
of $\alpha_{2000}$\,$=$\,17$^{h}$8$^{m}$10$^{s}$ and
$\delta_{2000}$\,$=$\,$-$44\degr21\arcmin, with a statistical error of
3\arcmin~on each axis, as indicated by the white cross in
Fig.~\ref{fig1}. Consequently, the new VHE $\gamma$-ray source is
called HESS\,J1708$-$443. The fit results also provide the intrinsic Gaussian
width, 0.29\degr\,$\pm$\,0.04\degr$_{\mathrm{stat}}$.

A preliminary differential energy spectrum was determined within a
circular region of 0.71\degr~radius (indicated by a dashed circle in
Fig.\,\ref{fig1}), chosen as a compromise between optimal
signal-to-noise ratio and independence of source morphology.
\begin{figure}[t!]
  \begin{center}
    \includegraphics[width=3.2in]{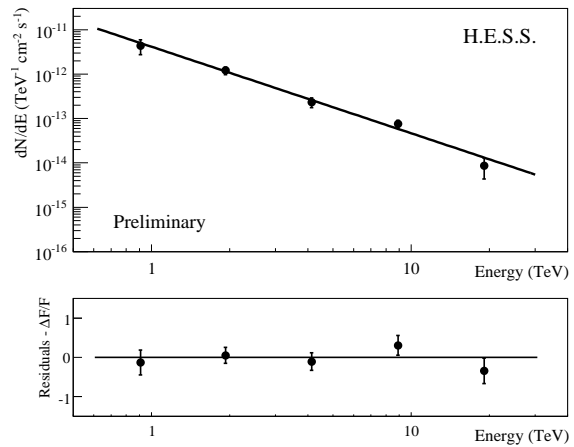}
    \caption[Differential energy spectrum of
    HESS\,J1708$-$443]{Differential energy spectrum of
      HESS\,J1708$-$443, extracted from the circular region indicated
      in Fig.~\ref{fig1}. The solid line shows the result of a 
      power law fit. The error bars denote 1-$\sigma$ statistical
      errors; the bottom panel shows the residuals of the fit.
      Events with energies between 600\,GeV and 28\,TeV were used
      in the determination of the spectrum.}
 \label{figSpectrum}
\end{center}
\end{figure}
Within this region, 605 excess events were found, corresponding to a
statistical significance of 6.7\,$\sigma$ (pre-trials). The spectrum
is well-described by a power law $\phi$\,$=$\,$\phi_{1\mathrm{TeV}}\cdot
\mathrm{E}^{-\Gamma}$ with a spectral index of
$\Gamma$\,$=$\,2.0\,$\pm$\,0.1$_{\mathrm{stat}}$\,$\pm$\,0.2$_{\mathrm{sys}}$
and a flux normalization at 1 TeV of
$\phi_{1\mathrm{TeV}}$\,$=$\,(4.2\,$\pm$\,0.8$_{\mathrm{stat}}$\,$\pm$\,
1.0$_{\mathrm{sys}}$)\,$\cdot\,
10^{-12}\,\mathrm{cm}^{-2}\mathrm{s}^{-1}\mathrm{TeV}^{-1}$. The
integral flux between 1 and 10\,TeV is $\sim$17\% of the Crab Nebula
flux in the same energy range. The flux points extracted from the
extended emission and the fitted power law are shown in
Fig.~\ref{figSpectrum}.

\section{The origin of the TeV emission}
While a superposition of multiple sources cannot be excluded, each of
the following objects could individually account for the observed VHE
$\gamma$-ray emission.

\subsection{A relic nebula from PSR\,B1706$-$44}
With its high spin-down luminosity of 3.4\,$\cdot\,
10^{36}$\,erg\,s$^{-1}$, the pulsar PSR\,B1706$-$44 is energetic enough
to power the observed VHE $\gamma$-ray emission. Assuming the pulsar is
at a distance of 2.5\,kpc, the energy flux from the H.E.S.S. source
between 1 and 10\,TeV is 1.2\,$\cdot$\,10$^{34}$\,erg\,s$^{-1}$. The
implied effective conversion efficiency from rotational
energy to $\gamma$-rays in this energy range is then $\sim$0.4\%,
comparable to the efficiency of 0.8\% inferred for PSR~J1420$-$6048
\cite{Aharonian_2006_Kookabura}. This suggests the pulsar's wind
nebula as a possible origin of the observed VHE $\gamma$-ray emission,
similar to other PWN associations such as Vela~X
\cite{Aharonian_2006_VelaX} and HESS\,J1825$-$137
\cite{Aharonian_2006_1825II}. In this scenario, the VHE
$\gamma$-emission originates from accelerated electrons which
up-scatter ambient photons to VHE energies (leptonic scenario).

The larger size of the TeV PWN compared to the ``bubble'' nebula seen
in X-rays (radius $\sim$110\arcsec) \cite{Romani_2005} can usually be
explained by the different energies, and hence cooling times, of the electrons
which emit X-rays and VHE $\gamma$-rays; such differences in
size have already been observed in other PWN associations such as
HESS\,J1825$-$137 \cite{Aharonian_2006_1825II}. However, in contrast
to the PWN of PSR~J1826$-$1334, where a magnetic field strength of
10\,$\mu$G was inferred from X-ray observations \cite{Gaensler_2003},
Romani et al.~\cite{Romani_2005} estimated a magnetic field as strong
as 140\,$\mu$G within the $\sim$\,110\arcsec~X-ray PWN of
PSR\,B1706$-$44. In such high magnetic fields, electrons that emit keV
X-rays and those that emit TeV $\gamma$-rays have comparable energies
and hence comparable cooling times. Thus, the TeV PWN should appear
almost point-like on the 5\arcmin~scale of the H.E.S.S. PSF.
Furthermore, given that the ratio of X-ray
to VHE $\gamma$-ray energy flux ($\mathrm{d}N/\mathrm{d}E\,\cdot\,E^2$) is determined by
the energy density in magnetic fields and inverse Compton (IC) target
photon fields (considering here only the CMB), the detected X-ray flux of
2.7\,$\cdot$\,10$^{-13}$\,erg\,cm$^{-2}$\,s$^{-1}$ at 1.7\,keV predicts a
$\gamma$-ray flux of 1.4\,$\cdot$\,10$^{-16}$\,erg\,cm$^{-2}$\,s$^{-1}$ at
1.7\,TeV, well below the level observable by H.E.S.S.

One way to reconcile the difference in emission region size 
and the high VHE flux level is to assume that the size
of the X-ray PWN is essentially governed by the extent of the
high-field region, and that the magnetic field falls off by a large
factor outside the X-ray PWN. The electrons can then escape from the
high-field region and---by accumulating over a significant fraction
of the pulsar's lifetime---form a larger nebula visible only in
VHE $\gamma$-rays.

This scenario still does not explain the asymmetry of the VHE
$\gamma$-ray nebula with respect to the pulsar location. Such
asymmetries have been observed before in other TeV PWNs,
e.g. HESS\,J1718$-$385, HESS\,J1809$-$193 \cite{Aharonian_2007_twopwn}
and HESS\,J1825$-$137
\cite{Aharonian_2005_1825I} \cite{Aharonian_2006_1825II}. They were
explained either by the proper motion of the pulsar or by a density
gradient within the ambient medium that either causes an asymmetry in
the reverse shock of the original supernova or different expansion
velocities of the TeV-emitting electrons
\cite{Blondin_2001} \cite{Swaluw_2001}.  In some of the simulations of
Swaluw et al.~\cite{Swaluw_2001}, the displaced PWN is indeed well-separated
from its pulsar. Both explanations are in principle applicable in this
situation. However, the measured scintillation velocity of less than
100\,km\,s$^{-1}$ for the pulsar renders the former explanation unlikely. The
latter explanation would favor a displacement of the TeV PWN towards a
low-density region, contrary to the observed offset, where the TeV emission
is closer to the higher density region along the Galactic plane. It
should be noted that a local density gradient, e.g. directly at the
position of the pulsar, could affect the spatial distribution of the
TeV PWN.

In this discussion it was assumed that the pulsar dominantly
accelerates electrons. If a considerable fraction of the accelerated
particles are hadrons, as discussed by Horns et al.~\cite{Horns_2007},
the constraints imposed by the large magnetic field within the X-ray
PWN are removed. The TeV emission would then originate from $\pi^0$
meson decay produced in inelastic interactions of accelerated protons
with ambient gas (hadronic scenario), and the VHE $\gamma$-ray
emission would trace the distribution of the target material. The
bright radio arc, which was interpreted by Bock et
al.~\cite{Bock_2002} as the compressed outer boundary of the former
wind-blown bubble, could act as such a region of enhanced target
material density, which would explain its coincidence with the
H.E.S.S. source.

\subsection{SNR G\,343.1$-$2.3}
The following discussion will investigate the scenario where the
VHE $\gamma$-ray emission originates in the SNR shell.  
The H.E.S.S. source is partially coincident with the bright radio arc
and the surrounding diffuse emission of the SNR, visible in the
1.4\,GHz observations taken with the ATCA instrument
\cite{Dodson_2002}. The best-fit position of the H.E.S.S. source is
consistent with the apparent center of the bright radio arc
($\alpha_{2000}$\,$=$\,17$^{h}$8$^{m}$ and
$\delta_{2000}$\,$=$\,$-$44\degr16\arcmin48\arcsec). However, due to relatively low
statistics in the VHE data, no further conclusions can be made about
morphological similarities.

Similar to the potential association with the PWN of PSR\,B1706$-$44,
both leptonic and hadronic scenarios for VHE $\gamma$-ray production
have to be considered. The leptonic scenario suffers from the
non-detection of the SNR at X-ray energies. The VHE $\gamma$-ray
spectrum reaches as far 20\,TeV. Assuming IC scattering
in the Thompson regime, the energy of the electrons upscattering CMB
photons up to 20\,TeV have an energy of roughly 80\,TeV. For a
reasonable magnetic field strength of 5\,$\mu$G, such electrons would
emit synchrotron photons with an energy of $\sim$1\,keV, i.e. within
the detectable energy range of current X-ray instruments. However, no
stringent UL on the X-ray flux from within the
H.E.S.S. source can be derived due to the vicinity of the luminous
low-mass X-ray binary 4U\,1705$-$440, whose stray light might be
obscuring diffuse X-ray emission from the SNR.

In the hadronic scenario, where synchrotron radiation is expected only
from secondary electrons, the lack of X-ray detection can easily be
accounted for. In this scenario, the total energy within the whole
proton population can be estimated by $W_{\mathrm{P}}(\mathrm{tot})
\,\approx\,3.9\,\times\,10^{49}\,\mathrm{erg}
\left(\frac{n}{\mathrm{cm}^{-3}}\right)^{-1}
\left(\frac{D}{\mathrm{kpc}}\right)^{2}$, following the
approach described in \cite{Hoppe_2007} (the proton spectrum was
assumed to follow a power law with a spectral index of $\alpha$\,=\,2
down to 1\,GeV). For a total energy of $10^{51}$\,erg released in the
supernova explosion, an acceleration efficiency of $\epsilon$\,=\,0.15
and a distance $D$\,$=$\,2.5\,kpc, the necessary average proton density is
$n$\,$\approx$\,1.6\,cm$^{-3}$, only slightly larger than the average
Galactic ambient density.

However, an association of SNR~G~343.1$-$2.3 with the pulsar
PSR\,B1706$-$44, a scenario debated in the literature, (see
e.g.~\cite{Bock_2002} and/or \cite{Romani_2005}), would make the SNR rather old (on
the order of 10\,000 yr) and place it in the late Sedov-Taylor phase, or
more likely, in the radiative phase. In this scenario, the SNR would
be older than SNRs from which shell-morphology $\gamma$-ray emission
has been unambiguously detected, e.g. RX\,J1713.7$-$3946
\cite{Aharonian_2007_RXJ1713} and RX\,J0852.0$-$4622
\cite{Aharonian_2007_VelaJr} ($\sim$2\,000\,yr).

\section{Summary}
H.E.S.S. observations have led to the discovery of a new VHE
$\gamma$-ray source, HESS\,J1708$-$443. The $\gamma$-ray emission is
extended, but the exact morphology of the emission region is still
under study. The flux from the source is $\sim$\,17\% of the Crab
Nebula flux, with a hard spectral index of 2.0. The possible
associations of HESS\,J1708$-$443 with a relic PWN of PSR\,B1706$-$44
and the SNR G\,343.1$-$2.3 have been discussed. Although a possible
association between the SNR and pulsar PSR\,B1706$-$44 suggests that
the SNR is in a later evolutionary stage than other
previously-detected VHE $\gamma$-ray emitting SNRs, there is at
present no ground to favor either of these two possible counterparts as
being associated with the H.E.S.S. source.

\section{Acknowledgments}
The support of the Namibian authorities and of the University of Namibia
in facilitating the construction and operation of H.E.S.S. is gratefully
acknowledged, as is the support by the German Ministry for Education and
Research (BMBF), the Max Planck Society, the French Ministry for Research,
the CNRS-IN2P3 and the Astroparticle Interdisciplinary Programme of the
CNRS, the U.K. Science and Technology Facilities Council (STFC),
the IPNP of the Charles University, the Polish Ministry of Science and 
Higher Education, the South African Department of
Science and Technology and National Research Foundation, and by the
University of Namibia. We appreciate the excellent work of the technical
support staff in Berlin, Durham, Hamburg, Heidelberg, Palaiseau, Paris,
Saclay, and in Namibia in the construction and operation of the
equipment.

\end{document}